\documentclass[aps,prb,twocolumn,showpacs,10pt,floatfix,superscriptaddress]{revtex4-1}
\usepackage[dvips]{graphics}
\usepackage{graphicx}
\usepackage[utf8]{inputenc}
\usepackage{amssymb}
\usepackage{epstopdf}
\usepackage{bm}
\usepackage{dcolumn}
\usepackage{epsfig}
\usepackage{latexsym}
\usepackage{amsmath}
\usepackage{color}
\usepackage{array}
\usepackage{enumerate,dsfont}

\def\XXint#1#2#3{{\setbox0=\hbox{$#1{#2#3}{\int}$ }
\vcenter{\hbox{$#2#3$ }}\kern-.5\wd0}}

\renewcommand{\vec}[1]{\bm{#1}}

\newcommand{\vctr}[1]{{\bf {#1}}}
\renewcommand{\vr}{\vctr{r}}

\newcommand{\PT}{T}
\newcommand{\lli}{l_{\rm a}}
\newcommand{\lle}{l_{\rm i}}

\makeatletter
\newcommand*{\rom}[1]{\expandafter\@slowromancap\romannumeral #1@}
\makeatother

\begin{document}

\title{Graphene {\em pn} junction in a quantizing magnetic field: Conductance at intermediate disorder strength}

\author{Christian Fr\"a{\ss}dorf, Luka Trifunovic, Nils Bogdanoff, and Piet W.\ Brouwer}
\affiliation{Dahlem Center for Complex Quantum Systems and, Institut f\"ur Theoretische Physik, Freie Universit\"at Berlin, Arnimallee 14, 14195 Berlin, Germany}
\date{\today}

\begin{abstract}
In a graphene {\em pn} junction at high magnetic field, unidirectional ``snake states'' are formed at the {\em pn} interface. In a clean {\em pn} junction, each snake state exists in one of the valleys of the graphene band structure, and the conductance of the junction as a whole is determined by microscopic details of the coupling between the snake states at the {\em pn} interface and quantum Hall edge states at the sample boundaries [Tworzydlo {\em et al.}, Phys. Rev. B {\bf 76}, 035411 (2007)]. Disorder mixes and couples the snake states. We here report a calculation of the full conductance distribution in the crossover between the clean limit and the strong disorder limit, in which the conductance distribution is given by random matrix theory [Abanin and Levitov, Science {\bf 317}, 641 (2007)]. Our calculation involves an exact solution of the relevant scaling equation for the scattering matrix, and the results are formulated in terms of parameters describing the microscopic disorder potential in bulk graphene.
\end{abstract}

\maketitle

\section{Introduction}
\label{sec:Introduction}

Many of the unique electronic properties of graphene, a single layer of carbon
atoms as they occur in graphite, can be traced back to its pseudorelativistic
band structure, in which quasiparticles behave as massless relativistic Dirac
particles, be it with the Fermi velocity $v_{\rm F}$ instead of the speed of
light $c$.~\cite{castroneto2009,dassarma2011,peres2010} Examples of such
``relativistic'' effects in graphene are Klein tunneling through potential
barriers,~\cite{klein1929,cheianov2006,katsnelson2006,beenakker2008} the
Zitterbewegung in confining potentials,\cite{katsnelson2006} the anomalous
integer quantum Hall
effect,~\cite{gusynin2005,novoselov2005,zhang2005,novoselov2007} or the
breakdown of Landau quantization in crossed electric and magnetic
fields.~\cite{lukose2007,peres2007}

The integer quantum Hall effect in graphene is called ``anomalous'' because the
number of chiral edge states at the boundary of a graphene flake in a large
perpendicular magnetic field is a multiple of four plus two, whereas the Dirac
bands are fourfold degenerate because of the combined spin and valley
degeneracies. The presence of a ``half'' edge mode per valley degree of freedom
has a direct explanation once it is taken into account that the valley
degeneracy is necessarily lifted at a graphene flake's outer
boundaries.~\cite{brey2006} Chiral states need not only occur at a flake's outer
boundaries, but they may also occur in the sample's interior, separating regions
with different electron density. At such an interface valley degeneracy is
usually preserved, and the number of chiral interface states is always a
multiple of four.

\begin{figure}
\includegraphics[width=0.8\columnwidth]{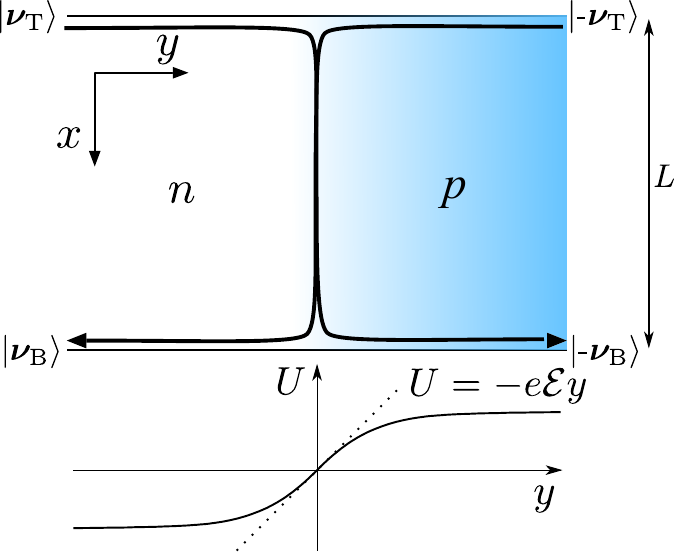}
\caption{(Color online)
Schematic experimental setup of a graphene {\em pn}-junction in a quantizing
magnetic field, such that the n region has filling fraction $2$ (left) and the
p region has filling fraction $-2$ (right). At the {\em pn} interface there is
a fourfold degenerate chiral interface state; there are twofold degenerate
chiral edge states at the sample's top and bottom edge.}
\label{fig:setup}
\end{figure}

A particularly interesting realization of such an interface occurs at a {\em
pn} junction in a perpendicular magnetic field, separating hole-doped (p-type)
and electron-doped (n-type) graphene
regions.~\cite{williams2007,lohmann2009,ki2009} The edge states at the {\em pn}
interface are referred to as ``snake states'' because, at least in a
semiclassical picture, such states propagate alternatingly at the $p$ and $n$
sides of the junction,~\cite{williams2011,rickhaus2015} similar to the behavior
of the states that propagate along zero-field contours in the quantum Hall
insulators in an inhomogeneous magnetic
field.~\cite{mueller1991,ye1995,reijniers2000,ghosh2008} A graphene {\em pn}
junction also has edge states at the sample boundaries, which move in opposite
directions in the {\em p} and {\em n}-type regions, see Fig.~\ref{fig:setup},
and feed into/flow out of the snake states at the {\em pn} interface.

The minimal number of chiral edge and interface states is realized for a {\em
pn} junction with filling fractions $2$ and $-2$. In this case there are two
edge modes, one for each spin direction, and four chiral interface modes. The
two-terminal conductance $G$ of such a {\em pn} junction is determined by the
probability $\PT$ that an electron that enters the common edge at the pn
interface from the source reservoir is transmitted to the drain
reservoir,
\begin{equation}
  G = \frac{2 e^2}{h} \PT \,.
\end{equation}
In the limit of a strongly disordered {\em pn} interface, Abanin and Levitov
predicted that the probability $\PT$ itself is subject to mesoscopic
fluctuations,~\cite{abanin2007} with average $\langle \PT \rangle = 1/2$ and
variance $\mbox{var}\, \PT = 1/12$.\footnote{Abanin and Levitov predict
  $\mbox{var}\, \PT = 1/15$ for $(\nu_n,\nu_p) = (2,-2)$ if spin-orbit coupling
  is strong enough that the spin degeneracy is lifted.~\cite{abanin2007} The
result quoted in the main text is valid in the presence of spin degeneracy.} In
the opposite limit of an ideal graphene sheet, Tworzydlo {\em et al.} found\cite{tworzydlo2007}
\begin{equation}
  \PT = \frac{1}{2}(1 - \vec{\nu}_{\rm T} \cdot \vec{\nu}_{\rm B}) \,,
\end{equation}
where the ``isospin'' vectors $\vec{\nu}_{\rm T}$ and $\vec{\nu}_{\rm B}$
describe the precise way in which the valley degeneracy is broken at the sample
boundaries, see Fig.~\ref{fig:setup}. Subsequent theoretical work involved a
semiclassical analysis,~\cite{carmier2010,carmier2011} numerical simulations of
the effect of disorder~\cite{li2008,long2008} and a phenomenological inclusion
of dephasing.~\cite{chen2011b} Several experimental groups have performed
measurements of the two-terminal conductance of graphene {\em pn} junctions in
a large perpendicular magnetic
field.~\cite{williams2007,lohmann2009,ki2009,ki2010,woszczyna2011,williams2011,schmidt2013,matsuo2015b,klimov2015}
The measured conductance follows the ensemble average of the strongly
disordered limit of Ref.\ \onlinecite{abanin2007}, although the experimentally
observed mesoscopic fluctuations remain significantly below the theoretical
prediction. Measurements of the shot noise power find a value that approaches
the theoretical prediction for the shortest interface
lengths.~\cite{matsuo2015,kumada2015}

In this article we present a theory of the transmission probability $\PT$ for a
graphene {\em pn} junction with generic disorder. We focus on the case of
filling fractions $(\nu_n,\nu_p) = (2,-2)$, for which we give an exact solution
for the distribution of the transmission probability $\PT$, thus bridging the
gap between the clean limit of Ref.\ \onlinecite{tworzydlo2007} and the
strong-disorder limit of Ref.\ \onlinecite{abanin2007}. Knowledge of the
distribution of $\PT$ allows us to calculate the average conductance $G$, its
variance, and the Fano factor $F$ throughout the weak-to-strong
disorder crossover. There are two reasons why we focus on the case
$(\nu_n,\nu_p) = (2,-2)$ for our exact solution. First, as we show below, two
length scales suffice to describe the effect of generic disorder on the edge
states, which is an essential simplification that makes our exact solution
possible. Second, quantum interference effects are strongest in this case, so
that the need for an exact treatment is maximal. Our results for the case
$(\nu_n,\nu_p) = (2,-2)$ also apply to higher filling fractions, if the mixing
of interface states occurs for the lowest Landau lavel only.~\cite{klimov2015}

The problem we consider here is related to two different problems that have
been studied in the literature, and we wish to comment on both. First, the
study is reminiscent of that of transport in coupled one-dimensional channels
with disorder, a problem that was solved exactly already in the 1950s, in the
context of wave propagation through random
media.~\cite{gertsenshtein1959a,gertsenshtein1959b} A crucial difference between the two problems is, however, that all one-dimensional modes at the {\em pn}
interface propagate in the same direction, whereas a normal metal wire has
equal numbers of modes propagating in both directions. This difference leads to
a rather different phenomenology: Whereas transmission is exponentially
suppressed for sufficiently strong disorder or long length in the standard
case,~\cite{mott1961} for the chiral interface states at a {\em pn} junction the
probability that electrons are transmitted along the interface is always one.
The question is whether they are fed into an edge state that transfers them
back to the source reservoir, or into the edge that leads to the drain.

The second related problem is that of the parametric dependence of transport
properties in mesoscopic samples. Traditionally (and correctly), it is the
Hamiltonian that is taken to depend on an external parameter, such as the
magnetic field or a gate voltage, either by modeling the perturbation directly,
or in a stochastic manner through a ``Brownian motion'' process. In a second
step the transport properties are then calculated from the Hamiltonian. There
have been theoretical attempts to make a theory directly for the parameter
dependence of the scattering matrix, {\em e.g.}, through a modification of
Dyson's Brownian motion model, but such an approach could not be made to agree
with the Hamiltonian-based approach if the dimension of the scattering matrix
is small.\cite{macedo1994,macedo1996,frahm1995c,rau1995} Interestingly, we find
that the dependence of the scattering matrix of the interface states on the
interface length is {\em precisely} described by the Dyson Brownian motion
model. To our knowledge, this constitutes the first application of this model
to a quantum transport problem.

The article is organized as follows. In Sec.~\ref{sec:MicroscopicModel} we
outline the microscopic model of a disordered graphene {\em pn}-junction and
derive an effective one-dimensional Hamiltonian for the chiral interface states
in the presence of generic disorder. In Sec.\ \ref{sec:conductance}, we then
derive and solve the Fokker-Planck equation describing the diffusive transport
through the {\em pn}-junction. Using the probability distribution of the
scattering matrix, we obtain an expression for the conductance and its
variance, being valid for an arbitrary disorder strengths. We conclude in Sec.\
\ref{sec:Conclusion}

\section{Microscopic Model}
\label{sec:MicroscopicModel}

We choose coordinates such that the {\em pn} interface is along the $x$ direction, see Fig.\ \ref{fig:setup}. At low energies conduction electrons in the graphene {\em pn} junction are described by a $4 \times 4$ matrix Hamiltonian,
\begin{equation}
  \hat H = \hat H_0 + \hat V(\vr) \,,
  \label{eq:H}
\end{equation}
in which $\hat V(\vr)$ in Eq.\ (\ref{eq:H}) is a matrix-valued potential representing the disorder and
\begin{equation}
  \hat H_0 = \tau_0 \otimes \sigma_0 U(y) + v_{\rm F} \tau_3 \otimes (\sigma_1 \pi_1(\vr) + \sigma_2 \pi_2(\vr)) \,.
  \label{eq:H0}
\end{equation}
Here the $\tau_{\mu}$ and $\sigma_{\mu}$ are Pauli matrices acting in valley and sublattice space, respectively, $U(y)$ is a gate potential that defines the {\em p} and {\em n}-type regions, and $\pi_1(\vr)$ and $\pi_2(\vr)$ are the in-plane components of the kinematic momentum,
\begin{align}
  \pi_1(\vr) &= -i \hbar \partial_x - e A_x(\vr) \,, \nonumber \\
  \pi_2(\vr) &= -i \hbar \partial_y - e A_y(\vr) \,.
  \label{eq:KineticMomentumOperators}
\end{align}
Since spin-orbit coupling is weak in graphene, the spin degree of freedom will be suppressed throughout.

For the vector potential we take the asymmetric gauge
\begin{equation}
  A_1(\vr) = - {\cal B} y \,,\ \ A_2(\vr) = 0 \,,
\end{equation}
with ${\cal B} > 0$ the perpendicular magnetic field. The magnetic field defines the length scale $\ell = (e {\cal B})^{-1/2}$. The gate potential $U(y)$ is negative for $y < 0$, zero for $y=0$, and positive for $y > 0$, so that the {\em pn} interface is at $y=0$ precisely, see Fig.\ \ref{fig:setup}. In the limit of a large magnetic field, it is sufficient to expand $U(y)$ to linear order in $y$ for $|y| \lesssim \ell$, and we set
\begin{equation}
  U(y) = - e {\cal E} y \,.
  \label{eq:Ulinear}
\end{equation}

In order to describe graphene with generic disorder we expand the matrix-valued disorder potential $\hat V(\vr)$ as\cite{aleiner2006,altland2006,ostrovsky2006}
\begin{equation}
  \hat{V}(\vec{r}) = \sum_{\mu, \nu = 0}^3 V_{\mu \nu}(\vec{r}) \tau_{\mu} \otimes \sigma_{\nu} \,,
\label{eq:DisorderMatrixExpansionTauSigmaBasis}
\end{equation}
with real amplitudes $V_{\mu \nu}(\vec{r})$. We assume these amplitudes to be Gaussian correlated with vanishing mean and with correlation function
\begin{align}
  \langle V_{\mu \nu}(\vec{r}) V_{\mu' \nu'}(\vec{r}') \rangle &= 
  \Gamma_{\mu \nu} \delta_{\mu \mu'} \delta_{\nu\nu'}
  \delta(\vec{r} - \vec{r}') \,,
\label{eq:StatisticsDisorderVariance}
\end{align}
where the absence of correlations between different amplitudes is a consequence of translation and rotation symmetry on the average.\cite{ostrovsky2006} The same symmetry considerations reduce the number of independent correlators to nine,
\begin{equation}
\Gamma_{\mu \nu} =
\begin{pmatrix}
	\alpha_0 & \gamma_{\perp} & \gamma_{\perp} & \alpha_z \\
	\beta_z & \beta_{\perp} & \beta_{\perp} & \beta_0 \\
	\beta_z & \beta_{\perp} & \beta_{\perp} & \beta_0 \\
	\gamma_0 & \alpha_{\perp} & \alpha_{\perp} & \gamma_z
	\end{pmatrix},
\label{eq:DisorderStrength}
\end{equation}
such that the five parameters $\alpha_0$, $\beta_{\perp}$, $\beta_z$, $\gamma_{\perp}$, and $\gamma_z$ represent disorder contributions respecting time-reversal-symmetry,\cite{aleiner2006,altland2006} whereas the remaining four parameters $\alpha_{\perp}$, $\alpha_z$, $\beta_0$, and $\gamma_0$ represent time-reversal-symmetry-breaking disorder. The coefficient $\alpha_0$ represents potential disorder that is smooth on the scale of the lattice spacing; the coefficients $\beta_{\perp}$ and $\gamma_z$ appear if the potential disorder is short range, so that it couples to the valley and sublattice degrees of freedom. The other coefficients are associated with a (random) magnetic field, strain, or lattice defects, see Ref.\ \onlinecite{ostrovsky2006}. Since time-reversal symmetry is broken by the large magnetic field ${\cal B}$, we will consider all nine contributions. 

With a large magnetic field ${\cal B}$ the low-energy degrees of freedom of the Hamiltonian (\ref{eq:H}) are the two chiral one-dimensional modes at the {\em pn} interface (per spin direction). They are described by an effective Hamiltonian
\begin{equation}
  H_{\rm s} = -i \hbar v_{\rm s} \tau_0 \partial_x + \sum_{\mu=0}^{3} V_{{\rm s},\mu}(x) \tau_{\mu} \,,
  \label{eq:Hs}
\end{equation}
where $v_{\rm s}$ is the velocity of the interface modes and the $V_{{\rm
s},\mu}(x)$ are effective disorder potentials representing the effect of the
bulk disorder potential $\hat V(\vr)$ on the interface states. In the limit of
a large magnetic field, we can find exact expressions for $v_{\rm s}$ and for
the correlation functions of the disorder potential $V_{\rm s}$ in terms of the
parameters of the underlying two-dimensional Hamiltonian (\ref{eq:H}). The
linear approximation (\ref{eq:Ulinear}) for the gate potential $U$ allows us to
make use of an exact solution for the eigenstates of the Hamiltonian $H_0$ of
Eq.\ (\ref{eq:H0}).\cite{peres2007,lukose2007} [See Ref.\ \onlinecite{liu2015}
for an approximate solution that does not make use of the linear approximation
(\ref{eq:Ulinear}).] Furthermore, for large magnetic fields the Landau level
separation is large enough that only the zeroth Landau level needs to be
considered. With the help of the exact solution for the zeroth Landau level we
then find that the velocity of the interface modes is
\begin{equation}
  v_{\rm s} = {\cal E}/{\cal B} \,,
  \label{eq:vs}
\end{equation}
whereas the disorder potentials $V_{{\rm s},\mu}(x)$ have zero mean and
correlation functions
\begin{align}
  \langle V_{{\rm s},\mu}(x) V_{{\rm s},\nu}(x') \rangle &= K_{\mu} 
  \delta_{\mu \nu} \delta(x - x') \,,
\label{eq:StatisticsEffectiveDisorder}
\end{align}
with, to leading order in $v_{\rm s}/v_{\rm F} \ll 1$,
\begin{subequations}
  \label{eq:Kmu}
\begin{align}
\label{eq:EffectiveDisorderStrengthsApproximatedalpha}
  K_0(\alpha_0,\alpha_z,\alpha_{\perp}) & = \frac{1}{\sqrt{2 \pi \ell^2}}  \left( \alpha_0 +  \alpha_z \right) \,, \\ 
  K_{1,2}(\beta_0,\beta_z,\beta_{\perp}) 
  &= \frac{1}{\sqrt{2 \pi \ell^2}}  \left( \beta_0 +  \beta_z \right) \,, \\
  K_3(\gamma_0,\gamma_z,\gamma_{\perp}) & = \frac{1}{\sqrt{2 \pi \ell^2}}  \left( \gamma_0 +  \gamma_z \right) \,.
\label{eq:EffectiveDisorderStrengthsApproximatedgamma}
\end{align}
\end{subequations}
The microscopic amplitudes $\alpha_{\perp}, \beta_{\perp}, \gamma_{\perp}$
contribute only at higher orders in $v_{\rm s}/v_{\rm F}$. We refer to
App.~\ref{sec:MappingFromMicroscopicToEffectiveModel} for details of the
calculation.

\section{Scaling approach for the scattering matrix}
\label{sec:conductance}
Disorder mixes the chiral interface modes. The effect of this disorder-induced
mode mixing is described by a $2 \times 2$ scattering matrix $\hat S$. In the
absence of disorder one has $\hat S = e^{ikL}\openone$. With disorder $\hat S$
acquires a nontrivial probability distribution $P(\hat S)$, which we now
calculate. 

We parametrize the scattering matrix using four ``angles'',
\begin{align}
  \hat S=e^{i\psi \tau_0}e^{i\tau_3 \varphi/2}e^{i\tau_2 \theta/2}e^{i\tau_3 \zeta/2} \,,
  \label{eq:Seuler}
\end{align}
where $\theta\in [0,\pi]$. We will first derive a differential equation that
describes the change of the joint distribution $P(\varphi,\theta,\zeta,\psi;L)$
upon changing the length $L$ of the interface region, see Fig.~\ref{fig:setup}.
To this end, we consider the scattering matrix $\hat S_{\delta L}$ for an
interface segment of length $\delta L$ much smaller than the mean free path for
disorder scattering. We parametrize $\hat S_{\delta L}$ as
\begin{equation}
  \hat S_{\delta L} = e^{i k \delta L} e^{i \hat A} \,,\ \
  \hat A = \sum_{\mu=0}^{3} r_{\mu} \tau_{\mu} \,.
  \label{eq:Aeuler}
\end{equation}
From the effective Hamiltonian (\ref{eq:Hs}) we find that the coefficients
$r_{\mu}$ are statistically independent, with disorder averages $\langle
r_{\mu} \rangle = 0$, $\mu=0,1,2,3$, and with variances
\begin{equation}
  \label{eq:rmuvar}
  \langle r_{\mu}^2 \rangle = \frac{K_{\mu}}{\hbar^2 v_{\rm s}^2} \delta L \,,
\end{equation}
with the coefficients $K_{\mu}$ given in Eq.~(\ref{eq:Kmu}). To simplify the
expressions in the remainder of this Section, we replace the notation with the
coefficients $K_{\mu}$ in favor of the inter-valley scattering length
\begin{equation}
  \lle = \frac{\hbar^2 v_{\rm s}^2}{4 K_1} \,,
\end{equation}
the (antisymmetric) intra-valley scattering length
\begin{equation}
  \lli = \frac{\hbar^2 v_{\rm s}^2}{4 K_3} \,,
\end{equation}
and the dimensionless coefficients 
\begin{equation}
  \alpha = K_0/4 K_1 \,, \ \ \gamma = K_3/K_1 = \lle/\lli \,,
\end{equation}
which relate inter- and intra-valley scattering rates. In the case of
pure potential disorder, only the disorder coefficients $\alpha_0$,
$\beta_{\perp}$, and $\gamma_z$ are nonzero, so that the constants $\alpha$,
$\gamma \sim (v_{\rm F}/v_{\rm s})^2 \gg 1$. For generic disorder that scatters
between the two sublattices of the hexagonal graphene lattice, one expects that
$\alpha$, $\gamma \sim 1$. The parameters $\alpha$ and $\gamma$ determine
symmetric and antisymmetric intra-valley scattering lengths, respectively. Since intra-valley
scattering that is equal for the two valleys corresponds to multiplication of
$\hat{S}$ with an overall phase factor, the coefficient $\alpha$ will not appear in
the expressions for the conductance distribution below. Antisymmetric
intravalley scattering, however, does affect the transmission probability $\PT$
of the {\em pn}~junction.

Since the interface modes are unidirectional, the composition rule for
scattering matrices is matrix multiplication. In particular, we obtain the
scattering matrix $\hat S(L + \delta L)$ of an interface segment of length $L +
\delta L$ as
\begin{equation}
  \hat S(L+ \delta L) = \hat S(L)\hat S_{\delta L} \,.
  \label{eq:Scomposition}
\end{equation}
This composition rule and the known statistical distribution of the scattering
matrices $\hat S_{\delta L}$ define a ``Brownian motion'' problem for the
scattering matrix $\hat S(L)$. An isotropic version of the Brownian motion
problem, with $\alpha = \gamma = 1$, was studied previously in the context of
quantum transport through chaotic quantum
dots.\cite{macedo1994,macedo1996,frahm1995c,rau1995} Using standard methods
(see App.\ \ref{sec:FP} for details), we can derive a Fokker-Planck equation
for the joint probability distribution $P(\varphi,\theta,\zeta,\psi;L)$ of the
coefficients parametrizing the scattering matrix $\hat S$,
\begin{align}
  \lle\frac{\partial P}{\partial L}=&-k\lle \frac{\partial P}{\partial\psi}+\frac{1}{2} \alpha
\frac{\partial^2 P}{\partial\psi^2}+\frac{1}{2} \left(\gamma+\cot^2\theta\right) 
    \frac{\partial^2 P}{\partial\zeta^2}\nonumber\\
    &+\frac{1}{2}\frac{\partial^2 P}{\partial \theta^2}-\frac{1}{2} \cot \theta  
    \frac{\partial P}{\partial \theta}+\frac{1}{2} \csc^2\theta 
    \frac{\partial^2 P}{\partial\varphi^2}\nonumber\\
    &-\cot \theta \csc\theta 
    \frac{\partial^2 P}{\partial\varphi\partial\zeta}+\frac{1}{2}\csc^2\theta P \,.
  \label{eq:FPeuler}
\end{align}
The Fokker-Planck equation Eq.~(\ref{eq:FPeuler}) for the $L$ dependence of the
scattering matrix of two co-propagating modes can be solved exactly by adapting
Ancliff's method to solve the corresponding problem for a pair of
counterpropagating modes.\cite{ancliff2016} After separating variables 
\begin{equation}
  P(L,\varphi,\theta,\zeta,\psi)=
  e^{-\lambda L/\lle}P(\varphi,\theta,\zeta,\psi) \,,
\end{equation}
Eq.\ (\ref{eq:FPeuler}) can be cast in the form of an eigenvalue problem,
which, following Ref.\ \onlinecite{ancliff2016}, can be solved exactly by
noticing that its right hand-side can be expressed through the operator $\hat
A$ defined in Eq.~(\ref{eq:Aeuler}), seen as a differential operator acting in
the Hilbert space of functions $f(\hat S)$,
\begin{equation}
  \langle \hat{A}^2\rangle  = - \left( \hat{L}_x^2 + \hat{L}_y^2 + \hat{L}_z^2 + (\gamma-1) \hat{L}_z^2 + \alpha \hat{L}_0^2 \right) \,,
  \label{eq:Alie}
\end{equation}
in which the operators $\hat{L}_{\mu}$ are the generators of the Lie algebra
$\mathfrak{u}(2)$. The Lie algebra $\mathfrak{u}(2)$ has two Casimir operators,
$\hat{L}_0$ and $\hat{\vec{L}}^2 = \hat{L}_x^2 + \hat{L}_y^2 + \hat{L}_z^2$,
that act as scalars $K$ and $l(l+1)$ ($l$ being integer or half-integer, $K$
being a real number), respectively, within each irreducible representation of
$U(2)$. Thus we can conclude immediately that the eigenvalues associated to the
eigenvalue problem obtained from Eq.~(\ref{eq:FPeuler}) are of the form
\begin{align}
  -\lambda_{Klm}&=l(l+1)+(\gamma-1)m^2+\alpha K^2+2ik\lle K,
  \label{eq:FPeigenvalues}
\end{align}
where $m = -l, -l + 1, \ldots, l$ and we included the drift term for $\psi$
being proportional to $k \lle$, which is not contained in Eq.\ (\ref{eq:Alie}).
The eigenfunctions can be expressed~\cite{landau1991} in terms of Jacobi
polynomials $P^{(a,b)}_n$ ($\vert m\vert\le l$)
\begin{align}
  \label{eq:FPeigenfunctions}
  P_{Klmn} &= \sqrt{\frac{(l+m)!(l-m)!}{(l+n)!(l-n)!}} e^{iK\psi} e^{im\varphi+in\zeta} \sin \theta \\
  &\times \sin^{m-n}(\theta/2) \cos^{m+n}(\theta/2) P_{j-m}^{(m-n,m+n)}(\cos\theta) \,. \nonumber
\end{align}
For $m=n=0$ these eigenfunctions match the ones previously obtained by Frahm
and Pichard for the isotropic scattering matrix Brownian motion
problem.~\cite{frahm1995c} It can be readily checked that the above functions
for arbitrary $K$, $l$, $m$, and $n$ are simultaneously eigenfunctions of
$\hat{\vec{L}}^2$, $\hat{L}_z$ and $\hat{L}_0$ and that they satisfy the
eigenvalue equation derived from Eq.~(\ref{eq:FPeuler}) with eigenvalues given
by Eq.~(\ref{eq:FPeigenvalues}). 

As the initial condition at $L=0$ we take $\hat S(0) = \openone$, which
corresponds to
\begin{equation}
  P(\varphi,\theta,\zeta,\psi;0) =
  \delta(\varphi+\zeta)\delta(\theta)\delta(\psi) \,.
\end{equation}
With this initial condition the solution for the probability distribution
is
\begin{align}
  P(\varphi,\theta,\zeta,\psi;L)=& \sqrt{\frac{\lle}{2 \pi \alpha L}}
  e^{-\frac{\lle (\psi-kL)^2}{2 \alpha L}}
  \sum_{l}\frac{2l+1}{8\pi^2} \sin \theta
  \nonumber \\ &\times
  \sum_{m = -l}^{l}
  e^{-[l(l+1)+(\gamma-1)m^2]L/\lle + i m(\varphi + \zeta)}\nonumber\\
  &\times
  \cos^{2m}(\theta/2)P_{l-m}^{(0,2m)}(\cos\theta) \,.
  \label{eq:FPsol}
\end{align}

The scattering matrix $\hat S$ is related to the transmission probability $\PT$
of a graphene {\em pn} junction through the relation~\cite{tworzydlo2007}
\begin{align}
  \PT &=\vert\langle\bm\nu_{\rm T}\vert \hat t_{\rm T}\hat S\hat t_{\rm B}\vert-\bm\nu_{\rm B}\rangle\vert^2 \,,
  \label{eq:Teh}
\end{align}
in which $\hat t_{\rm T}$ ($\hat t_{\rm B}$) is the scattering matrix
describing how the edge modes at the top (bottom) edges of the {\em pn}
junction feed into/originate from the interface modes and $\vert \pm \bm\nu_{T}
\rangle$ ($\vert \pm \bm\nu_{B} \rangle$) are valley isospin Bloch vectors for
the top (bottom) edges of the {\em n} ($+$) and {\em p}-doped ($-$) regions,
see Fig.\ \ref{fig:setup}. The isospin vectors $\vert \bm\nu_{X} \rangle$ are
superpositions of the vectors $\vert 1 \rangle$ and $\vert -1 \rangle$
representing the two valleys,
\begin{align}
  \vert\bm\nu_X\rangle=&
  \cos\frac{\theta_X}{2}\vert 1\rangle+e^{i\phi_X}
  \sin\frac{\theta_X}{2}\vert -1 \rangle \,, \nonumber \\ 
  \vert-\bm\nu_X\rangle=&
  \sin\frac{\theta_X}{2}\vert 1\rangle-e^{i\phi_X}
  \cos\frac{\theta_X}{2}\vert -1\rangle \,,
\end{align}
with polar angles $\theta_X$ and $\phi_X$, $X={\rm T}$, ${\rm B}$. The scattering matrices
$\hat t_{\rm T}$ and $\hat t_{\rm B}$ express isospin conservation at the point
where the valley-non-degenerate edge states merge into/evolve out of the valley
degenerate interface state,~\cite{tworzydlo2007}
\begin{equation}
  \hat t_X = 
  e^{i\tilde\varphi_X}\vert\bm\nu_X\rangle\langle\bm\nu_X\vert+e^{i\tilde\varphi_X^\prime}\vert-\bm\nu_X\rangle\langle-\bm\nu_X\vert \,,
  \label{eq:tLR}
\end{equation}
with $\tilde\varphi_X$ and $\tilde\varphi_X^\prime$ arbitrary phases that do
not need to be specified. Combination of Eqs.~(\ref{eq:Teh}) and (\ref{eq:tLR})
gives\cite{tworzydlo2007}
\begin{align}
  \PT &= \vert\langle\bm\nu_{\rm T}\vert \hat S\vert-\bm\nu_{\rm B}\rangle\vert^2 \,.
  \label{eq:TehS}
\end{align}
Using Eq.~(\ref{eq:Seuler}) as well as the fact that the phase difference
$\varphi-\zeta$ is uniformly distributed for all $L$, we find that the disorder
average $\langle \PT \rangle$ is given by
\begin{align}
  \label{eq:barTeh}
  \langle \PT \rangle =&\frac{1}{2}\Big[ 1 - \cos \theta_{\rm T} \cos \theta_{\rm B} \langle \cos \theta \rangle \\
    &- \sin \theta_{\rm T} \sin \theta_{\rm B} \langle \cos \theta \cos(\varphi+\phi_{\rm T}) \cos(\zeta-\phi_{\rm B}) \rangle \nonumber \\
    &+ \sin \theta_{\rm T} \sin \theta_{\rm B} \langle \sin(\varphi + \phi_{\rm T}) \sin(\zeta - \phi_{\rm B}) \rangle \Big] \,. \nonumber
\end{align}
Using the probability distribution (\ref{eq:FPsol}) one then finds the remarkably simple result
%
%
\begin{align}
  \langle \PT \rangle = &\frac{1}{2} \Big[ 1 - e^{-2L/\lle} \cos \theta_{\rm T} \cos \theta_{\rm B} \nonumber \\
  &- e^{-L/\lle - L/\lli} \sin \theta_{\rm T} \sin \theta_{\rm B} \cos(\phi_{\rm T}-\phi_{\rm B}) \Big] \,.
  \label{eq:Pavg}
\end{align}
Similarly we obtain the variance of the transmission probability
\begin{align}
\label{eq:varT}
\text{var}\, \PT&=\frac{1}{12}-\frac{1}{4} e^{-4L/\lle} \cos^2\theta_{\rm T}
  \cos^2\theta_{\rm B}\\
  &+\frac{1}{24} e^{-6L/\lle} (3 \cos^2\theta_{\rm T}-1) (3\cos^2\theta_{\rm B}-1) \nonumber \\
  &-\frac{1}{4} e^{-2 L/\lle - 2 L/\lli} \cos^2(\phi_{\rm T}-\phi_{\rm B})
  \sin^2\theta_{\rm T} \sin^2\theta_{\rm B} \nonumber \\
  &+\frac{1}{8} e^{-2L/\lle - 4 L/\lli} \cos 2 (\phi_{\rm T}-\phi_{\rm B})
  \sin^2\theta_{\rm T} \sin^2\theta_{\rm B} \nonumber \\
  &+\frac{1}{8} e^{-5 L/\lle - L/\lli} \cos(\phi_{\rm T}-\phi_{\rm B}) \sin(2
  \theta_{\rm T}) \sin(2 \theta_{\rm B}) \nonumber \\
  &-\frac{1}{8} e^{-3L/\lle - L/\lli}
  \cos(\phi_{\rm T}-\phi_{\rm B}) \sin(2 \theta_{\rm T}) \sin(2 \theta_{\rm B}) \,. \nonumber
\end{align}
In the isotropic case, $\gamma=\lle/\lli=1$, these expressions can be further
simplified, such that $\langle \PT \rangle$ and $\mbox{var}\, \PT$ depend on
the scalar product $\bm\nu_{\rm T}\cdot\bm\nu_{\rm B}$ of the isospin vectors
only,
\begin{align}
    \langle \PT \rangle = &\frac{1}{2} \left(1-e^{-2L/\lle}\bm\nu_{\rm T}\cdot\bm\nu_{\rm B} \right) \,, \\
    \text{var}\, \PT = &\frac{1}{12} - \frac{1}{4} e^{-4L/\lle} (\bm\nu_{\rm T}\cdot\bm\nu_{\rm B})^2 \nonumber \\
    &+\frac{1}{4}e^{-6L/\lle} \left((\bm\nu_{\rm T}\cdot\bm\nu_{\rm B})^2-\frac{1}{3} \right) \,,
  \label{eq:Tehlimiso}
\end{align}
In the limiting cases $L \ll \lle$, $\lli$ and $L \gg \lle$, $\lli$
Eqs.~(\ref{eq:Pavg}) and (\ref{eq:varT}) [or (\ref{eq:Tehlimiso})] agree with
the known results for the clean and dirty limits, respectively, see
Refs.~\onlinecite{abanin2007} and \onlinecite{tworzydlo2007}.

\begin{figure}
\includegraphics[width=0.8\columnwidth]{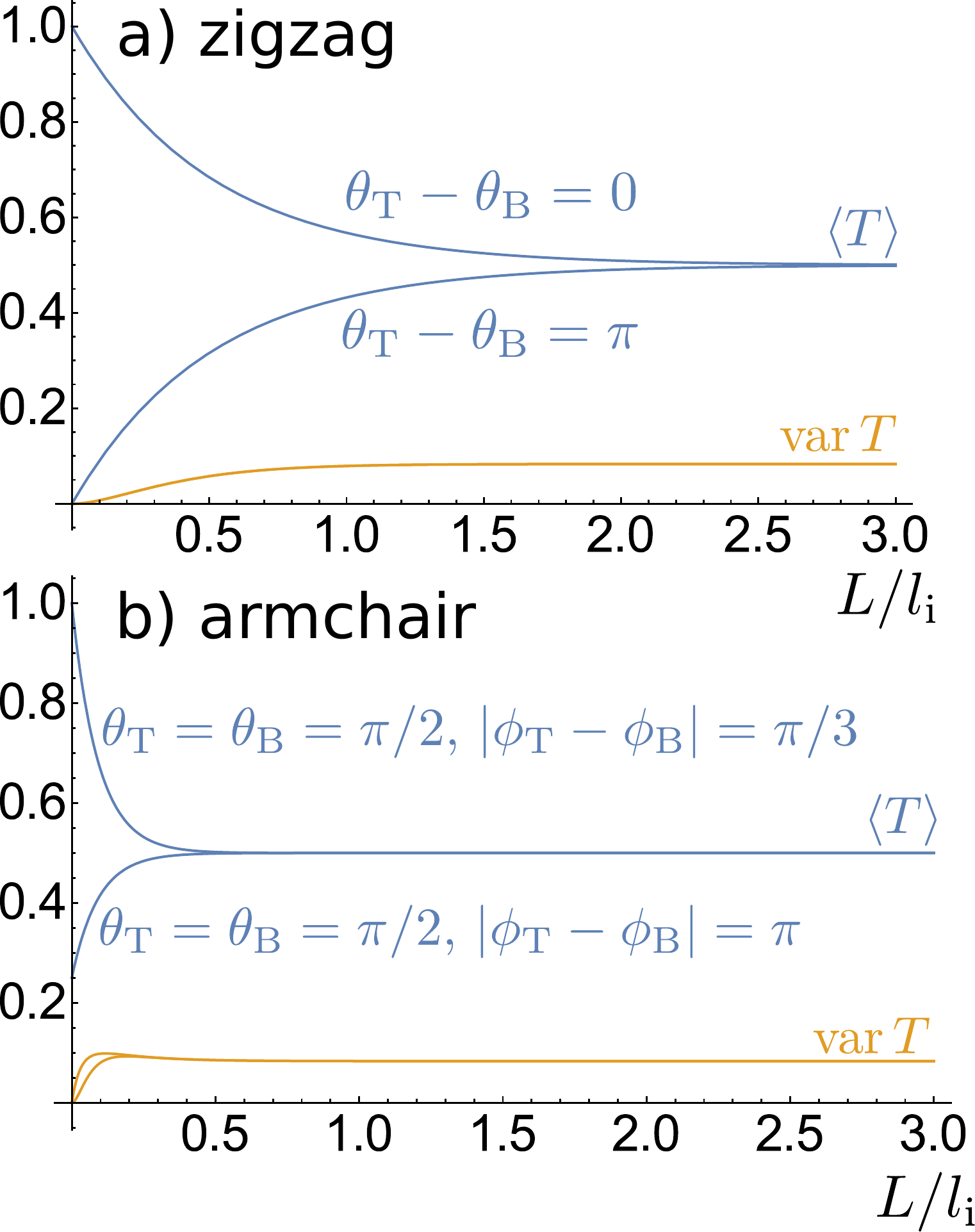}
\caption{(Color online) Mean $\langle\PT\rangle$ and variance $\mbox{var}\,\PT$
of the transmission $T$, as a function of the interface length $L$, for $\gamma
= \lle/\lli = 10$. Panel a shows results for zigzag termination of the
hexagonal lattice at the top and bottom edges; panel b is for armchair
termination. The top curve for $\mbox{var}\, \PT$ in panel b is for
$|\phi_{\rm T} - \phi_{\rm B}| = \pi/3$; the bottom variance curve is for
$|\phi_{\rm T} - \phi_{\rm B}| = \pi$.}
\label{fig:TvarT}
\end{figure}

Figure \ref{fig:TvarT} shows the ensemble average $\langle T \rangle$ and the
variance $\mbox{var}\, T$ for representative lattice terminations at the top
and bottom edges of the {\em pn} junction. For zigzag termination, one has
$|\bm{\nu}_X \cdot \bm{e}_z|=1$, so that the difference $\theta_{\rm T} -
\theta_{\rm B} = 0$ or $\pi$. Which of these two values is taken depends on the
parity of the number of hexagons along the interface length
$L$.\cite{tworzydlo2007} For armchair termination one has $\bm\nu_X \cdot
\bm{e}_z = 0$, so that $\theta_{\rm T} = \theta_{\rm B} = 0$. The difference
$\phi_{\rm T} - \phi_{\rm B}$ of the azimuthal angles can take the three values
$\pi$ and $\pm \pi/3$, depending on the number of hexagons along the interface
length $L$ modulo $3$. For the zigzag nanoribbon termination, the crossover
between the clean and strong disorder limits shows that the approach to the
average value and the development of large mesoscopic fluctuations occur at the
length scale $\lle$, whereas the characteristic length scale for armchair
nanoribbon termination is $\lli$.

Additional information on the mixing of interface states can be obtained from a
measurement of the Fano factor $F = P/2 e I$, the ratio of the shot noise power
$P$ and the current $I$. For the case we consider here, one has (at zero
temperature) \cite{blanter2000b} 
\begin{equation}
  F = 1-T \,,
\end{equation}
so that the ensemble average of the Fano factor $F$ directly follows from our
expression Eq.~(\ref{eq:Pavg}) for the disorder averaged transmission
probability $\PT$. In particular, in the limit of a clean junction ($L \ll
\lle$, $\lli$), one finds $F =(1+ \bm \nu_{\rm T} \cdot \bm \nu_{\rm
B})/2$, whereas in the limit of a dirty junction one has 
\begin{equation}
  \langle F \rangle = 1/2 \,.
  \label{eq:F1}
\end{equation} 

A finite temperature leads, first and foremost, to a smearing of the electron
energy. Since thermal smearing effectively amounts to taking an ensemble
average, thermal smearing has no effect on the ensemble average $\langle T
\rangle$, but it strongly suppresses the transmission fluctuations. In the
limit of large temperatures ($k_{\rm B} T$ much larger than the Thouless energy
of the interface) the Fano factor becomes\cite{blanter2000b} $F = \langle
T(1-T) \rangle/\langle T \rangle$, which may be easily evaluated by combining
Eqs.\ (\ref{eq:Pavg}) and (\ref{eq:varT}). In the limit of a clean junction one
then finds the same Fano factor as in the zero temperature limit, whereas in
the strong disorder limit $L \gg \lle$, $\lli$ the high-temperature limit is 
\begin{equation}
  \langle F \rangle = 1/3 \,.
  \label{eq:F2}
\end{equation}
Note that this value for $\langle F \rangle$, as well as the zero-temperature
limit (\ref{eq:F1}) mentioned above, differ from the Fano factor
reported in Ref.\ \onlinecite{abanin2007}. The difference arises, because Ref.\
\onlinecite{abanin2007} takes the semiclassical expression for the shot noise
power, whereas quantum effects are strong in the limit of low filling fractions
we consider here and the semiclassical approximation is no longer
quantitatively correct.

\begin{figure}
\includegraphics[width=0.8\columnwidth]{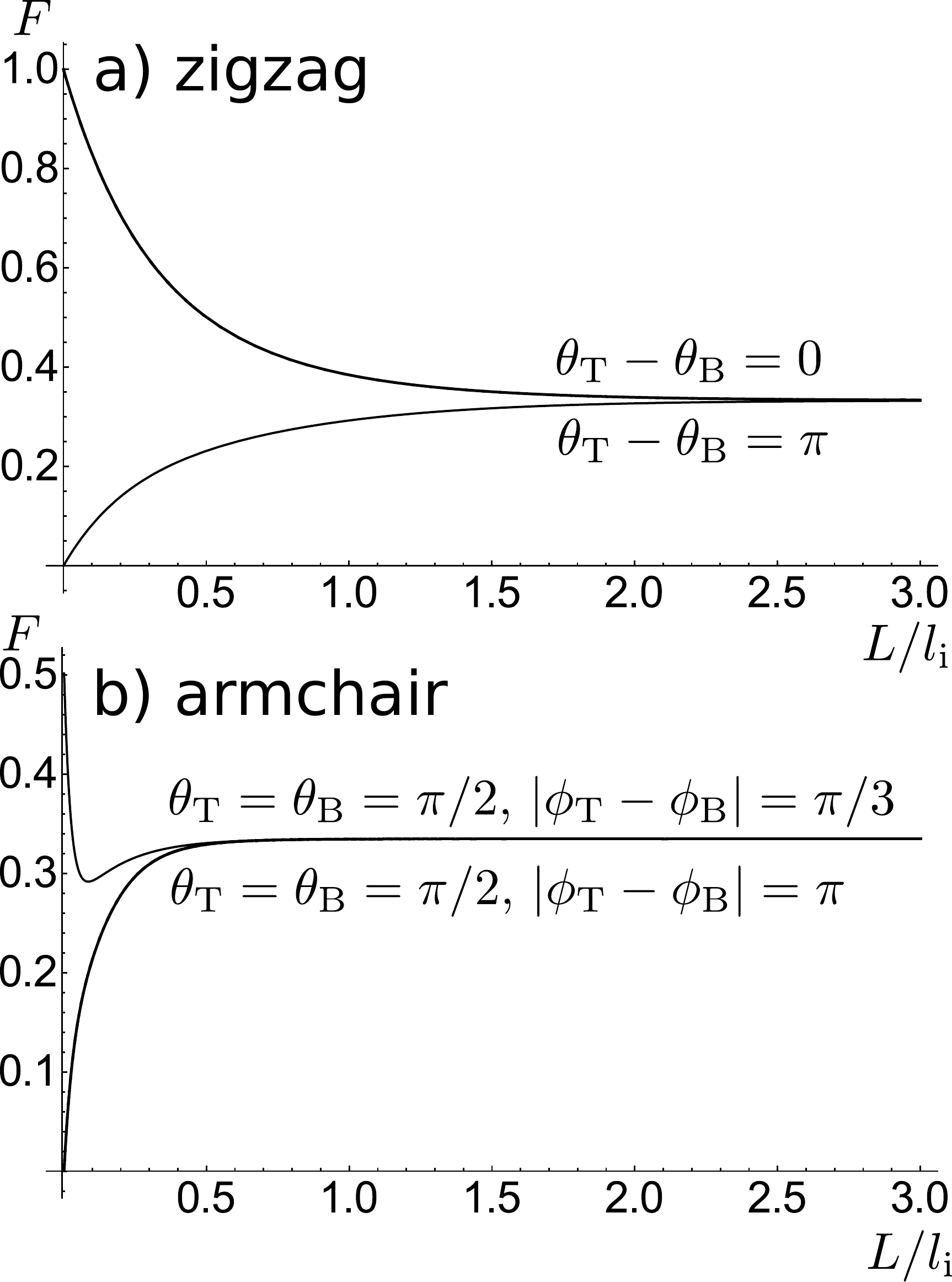}
\caption{The Fano factor $F$ versus interface length $L$ in the  high-temperature limit for $\gamma = \lle/\lli = 10$ and zigzag termination of the top and bottom edges (panel a) or armchair termination of the top and bottom edges (panel b).}
\label{fig:Fano}
\end{figure}

Figure~\ref{fig:Fano} shows the high-temperature limit of the Fano factor $F$
for the same representative edge terminations as in Fig.\ \ref{fig:TvarT}. For
the zigzag termination of top and bottom edges, the Fano factor monotonously
appraoches the large-$L$ asymptote (\ref{eq:F2}), with characteristic length
scale $\lle$. For armchair termination the dependence can be non-monotonic, and
the characteristic length scale is $\lli$. In the isotropic limit $\gamma =
\lle/\lli = 1$ both termination types exhibit a monotonous dependence on $L$
(data not shown). 

%
%
%
%
%

\section{Conclusion}
\label{sec:Conclusion}
We calculated the conductance distribution of a graphene {\em pn} junction in a
quantizing magnetic field. Our theory captures the entire crossover between the
limit of a clean {\em pn} junction and that of a strongly disordered junction.
In the former case, the conductance is a known function of the isospin vectors
$| \bm\nu_{T}\rangle$ and $|\bm\nu_{B} \rangle$ for the chiral states at the
edges of the {\em pn} junction.~\cite{tworzydlo2007} In the latter case the
conductance has a probability distribution that is universal and independent of
the details of the edges.~\cite{abanin2007} Our solution for the intermediate regime combines features of both
extremes: On the one hand, the conductance has finite sample-to-sample
fluctuations, on the other hand mean and variance of the conductance depend on
the isospin vectors $| \bm\nu_{T}\rangle$ and $|\bm\nu_{B} \rangle$.

A special feature of our solution is that we are able to relate the mean free
paths for transport along the one-dimensional interface to the coefficients
describing the random potential in the two-dimensional graphene sheet. Even
after translation and rotation invariance are taken into account, generic
disorder in graphene is still characterized by five independent constants. Some
information on these constants can be obtained from a measurements of a
two-dimensional graphene sheet. For example, pure potential disorder gives rise
to weak antilocalization, whereas disorder terms that couple the valleys cause
weak localization.~\cite{mccann2006,morpurgo2006,tikhonenko2009} Complementary information can be obtained from the carrier-density dependence of the conductivity.\cite{zhao2015} Our theory
links the conductance distribution of a {\em pn} junction in a large magnetic
field to the same set of coefficients and, thus, provides an additional and
independent method to determine these.

A central observation of the many conductance
experiments~\cite{williams2007,lohmann2009,ki2009,ki2010,woszczyna2011,williams2011,schmidt2013,matsuo2015b,klimov2015}
is that the measured conductance in the case $(\nu_n,\nu_p) = (2,-2)$
consistently agrees with the ensemble average $\langle \PT \rangle = 1/2$ of
the strong disorder limit,~\cite{abanin2007} but the experiments do not show any
signatures of the large mesoscopic fluctuations that are expected in the limit
of zero temperature. These experiments are not consistent with the clean-limit
predictions, since none of the standard nanoribbon terminations (armchair or
zigzag) gives a conductance $G$ consistent with $\PT=1/2$.~\cite{tworzydlo2007}
The Fano factors observed in Refs.~\onlinecite{kumada2015} and
\onlinecite{matsuo2015} are slightly below the theoretical predictions
Eqs.~(\ref{eq:F1}) and (\ref{eq:F2}) for the strong disorder limit (assuming spin
degeneracy), but not far from it when extrapolating the observation of
Ref.~\onlinecite{kumada2015} to zero interface length. Our theory for the
crossover between the clean and strong disorder limits shows that the approach
to the average value $\PT=1/2$ and the development of large mesoscopic
fluctuations occur at the same length scale, $\lle$ ($\lli$) for zigzag
(armchair) nanoribbon termination, irrespective of the form of the microscopic
disorder, see Fig.~\ref{fig:TvarT}. We note that while for non-standard
nanoribbon termination with $\vert\phi_T-\phi_B\vert=\pi/2$, it is possible to
approach the mean value $\PT=1/2$ on length scale $\lle$ while the mesoscopic
fluctuations are developed on the length scale $\lli$. The opposite
scenario, which would offer an explanation for the experimental observations,
is not possible within our theory. 
Other causes of a suppressed mesoscopic fluctuations
that have been mentioned in the literature are thermal smearing, slow time-dependent fluctuations
of system parameters, or inelastic processes contribution to the mixing between
the interface states.~\cite{abanin2007} The observed suppression of shot noise
for long interface lengths in Ref.~\onlinecite{kumada2015} clearly hints at a
role of inelastic processes for large interface lengths $L$, whereas the
observation of a finite shot noise power at shot junction lengths is consistent
with the first two explanations. A quantitative theory of thermal smearing
effects requires the extension of the present theory to the energy dependence of
the scattering matrix, a considerable theoretical challenge that is left to
future work.

\acknowledgments
This work is supported by the German Research
Foundation (DFG) in the framework of the Priority Program 1459 ``Graphene''.

\appendix
\section{Effective Hamiltonian for chiral interface states}
\label{sec:MappingFromMicroscopicToEffectiveModel}
In this appendix we derive the effective one-dimensional Hamiltonian $H_{\rm s}$
for the chiral states at the {\em pn} interface, see Eq.\ (\ref{eq:Hs}). Hereto
we need the explicit form of the eigenfunctions of the Hamiltonian $H_0$ for the
clean system. These eigenfunctions are known from the exact solution of Refs.\
\onlinecite{lukose2007,peres2007}. They have a linear energy-momentum dispersion
$\varepsilon_k = v_{\rm s} k$ with $v_{\rm s}$ given by Eq.\ (\ref{eq:vs}), and
the delta-function normalized spinor-valued wavefunctions for the zeroth Landau
level read\cite{lukose2007,peres2007}
\begin{equation}
  |\Psi_{k \kappa}^0(\vec{r})\rangle = e^{i k x} \phi_0 ( y - k \ell^2) |\kappa\rangle \otimes |\xi_{\kappa}\rangle \,,
\label{eq:ZeroModeSolutions}
\end{equation}
where $\kappa = \pm 1$ is the valley index, $|\kappa\rangle$ are the basis spinors with respect to the valley degree of freedom, and $|\xi_{\kappa}\rangle$ represents a two-component spinor with respect to the sublattice degree of freedom. Further, 
\begin{equation}
  \phi_0 \left( y \right) = \left( \frac{\beta}{\pi \ell^2} \right)^{1/4}
  e^{- \beta y^2/2 \ell^2} \,,
\label{eq:ZeroLandauLevel}
\end{equation}
where we abbreviated
\begin{equation}
  \beta = \sqrt{1 - \left( \frac{\mathcal{E}}{v_F \mathcal{B}} \right)^2},
\end{equation}
(Note that the validity of this exact solution requires $|{\cal E}| < v_{\rm F} {\cal B}$.) The spinor $|\xi_{\kappa}\rangle$ reads
\begin{align}
  \vec{\xi}_{\kappa} \equiv \sqrt{\frac{|\mathcal{E}|}{2 v_F \mathcal{B}}} \begin{pmatrix} \textrm{sign}(\mathcal{E}) \kappa C^{1/2} \\ C^{-1/2} \end{pmatrix} \,.
\label{eq:DefinitionXi}
\end{align}
with
\begin{align}
C &= \frac{v_F \mathcal{B}}{|\bar{\mathcal{E}}|} \left( 1 - \beta \right) \,.
\label{eq:DefinitionC}
\end{align}
One verifies that in the limit of vanishing electric field the solutions Eq.~\eqref{eq:ZeroModeSolutions} reduce to the well-known results for graphene in a homogeneous external magnetic field. 

As explained in the main text, for large magnetic fields it is sufficient to restrict to the zeroth Landau level. We may obtain an effective Hamiltonian for the interface states by projecting the Hamiltonian $H_0$ to the states spanned by the wavefunctions (\ref{eq:ZeroModeSolutions}). Using the Fourier representation of Eq.\ (\ref{eq:ZeroModeSolutions}) this projection takes the simple diagonal form
\begin{equation}
  H_{{\rm s},0} = v_{\rm s} k \tau_0 \,.
\end{equation}
Fourier transformation with respect to $k$ gives the first term of the Hamiltonian $H_{\rm s}$ of Eq.\ (\ref{eq:Hs}).

To incorporate the disorder potential we need to evaluate the matrix elements
\begin{align}
  V_{{\rm s},\kappa \kappa'}(k,k') = &\int d\vec{r} \langle \Psi_{k \kappa}^0(\vec{r})| \hat{V}(\vec{r}) | \Psi_{k'\kappa'}^0(\vec{r}) \rangle \nonumber \\
= &\int d\vec{r} e^{-i (k - k') x} 
  \phi_0(y - k \ell^2) \phi_0(y- k' \ell^2)
  \nonumber \\ & \times
  (\langle \kappa | \otimes \langle \xi_{\kappa} |) \hat V(\vr)
  (|\kappa'\rangle \otimes |\xi_{\kappa'}\rangle) \,.
\label{eq:EffectiveDisorderAppendix}
\end{align}
In the limit of a large magnetic field and for small momenta $k$, $k'$, we may neglect the shifts $k \ell^2$ and $k' \ell^2$ in the arguments of the functions $\phi_0$. With this approximation, $V_{{\rm s},\kappa \kappa'}(k,k')$ becomes a function of the difference $k-k'$ only, so that it represents an effective disorder potential that is local in space,
\begin{align}
  V_{{\rm s},\kappa \kappa'}(x) = \int dy \phi_0(y)^2
  (\langle \kappa | \otimes \langle \xi_{\kappa} |) \hat V(x,y)
  (|\kappa'\rangle \otimes |\xi_{\kappa'}\rangle).
  \label{eq:LocalOneDimensionalImpurityPotentials}
\end{align}
{}Since the disorder potential $\hat V(x,y)$ has a Gaussian distribution with zero mean and with delta-function correlations, the same applies to the effective disorder potential $\hat V_{\rm s}(x)$ for the interface states. The two-point correlation function can be calculated with the help of Eq.\ (\ref{eq:StatisticsDisorderVariance}), and one finds
\begin{equation}
  \langle V_{{\rm s},\kappa \lambda}(x) V_{{\rm s},\kappa' \lambda'}(x') \rangle = K_{\kappa \lambda \kappa' \lambda'} \delta(x - x') \,,
\label{eq:StatisticsEffectiveDisorderFlavourBasis}
\end{equation}
with
%
\begin{align}
K_{++++} &= K_{----} \equiv K_0 + K_3 \,,
\nonumber \\
K_{++--} &= K_{--++} \equiv K_0 - K_3 \,,
\label{eq:DisorderStrengths} \\
K_{+--+} &= K_{-++-}  \equiv 2 K_1 \,, \nonumber
\
\end{align}
%
where the coefficients $K_{\mu}$ are
\begin{subequations}
  \label{eq:Kmuapp}
\begin{align}
K_0 =&\frac{1}{4} \sqrt{\frac{\beta}{2 \pi  \ell^2}} \left( \frac{\mathcal{E}}{v_F \mathcal{B}} \right)^2 \Big( (C + 1/C)^2 \alpha_0 \nonumber \\
&+ (C  - 1/C)^2 \alpha_z + 4 \alpha_{\perp} \Big) \,,
\label{eq:DisorderStrengthMicroscopicCalculationAlphaS}
\\
  K_1 = K_2 =&\frac{1}{4} \sqrt{\frac{\beta}{2 \pi  \ell^2}} \left( \frac{\mathcal{E}}{v_F \mathcal{B}} \right)^2 \Big( (C + 1/C)^2 \beta_0 \nonumber \\
&+ (C  - 1/C)^2 \beta_z + 4 \beta_{\perp} \Big) \,,
\label{eq:DisorderStrengthMicroscopicCalculationBetaS}
\\
  K_3=&\frac{1}{4} \sqrt{\frac{\beta}{2 \pi  \ell^2}} \left( \frac{\mathcal{E}}{v_F \mathcal{B}} \right)^2 \Big( (C + 1/C)^2 \gamma_0 \nonumber \\
&+ (C  - 1/C)^2 \gamma_z + 4 \gamma_{\perp} \Big) \,.
\label{eq:DisorderStrengthMicroscopicCalculationGammaS}
\end{align}
\end{subequations}
Notice that each of the three coefficients depends on a different set of the disorder coefficients for the two-dimensional disorder potential $\hat V(x,y)$. Upon writing
\begin{equation}
  \hat V_{\rm s}(x) = \sum_{\mu=0}^{3} V_{{\rm s},\mu}(x) \tau_{\mu},
\end{equation}
the correlation function of the form (\ref{eq:StatisticsEffectiveDisorderFlavourBasis}) reproduces that of Eq.\ (\ref{eq:StatisticsEffectiveDisorder}) of the main text. The expressions for the coefficients $K_{\mu}$ quoted in Eq.\ (\ref{eq:Kmu}) of the main text follow from Eq.\ (\ref{eq:Kmuapp}) upon keeping the leading contribution in $({\cal E}/v_{\rm F} {\cal B})^2$.

\section{Derivation of the Fokker-Planck equation for scattering matrix}
\label{sec:FP}
In this appendix we give the details of the derivation of the Fokker-Planck
equation, Eq.~(\ref{eq:FPeuler}). We use the parameterization (\ref{eq:Seuler})
of the scattering matrix in terms of Euler angles, which we combine into a
four-component vector $\bm p=(\varphi,\theta,\zeta,\psi)^T$. The composition
rule (\ref{eq:Scomposition}) leads to a Langevin process for the Euler angles
$\bm p$. We can calculate the change $\delta \bm p$ from the change
\begin{equation}
  \delta \hat S = \hat S(L + \delta L) - \hat S(L)
\end{equation}
of the scattering matrix. We keep contributions to $\delta \bm p$ and $\delta
\hat S$ up to second order in $r_{\mu}$ and write accordingly
\begin{align}
  \delta \bm p =& \delta\bm p^{(1)}+\delta\bm p^{(2)}, \nonumber \\
  \delta \hat S =& \delta \hat S^{(1)}+\delta \hat S^{(2)}+O(r_{\mu}^3).
\end{align}
We can then obtain $\delta \bm p$ from $\delta \hat S$ using the relations
\begin{align}
  \delta \hat S^{(1)}&=\sum_{\mu=0}^3\frac{\partial \hat S}{\partial p_{\mu}}\delta p_{\mu}^{(1)} \,,\\
  \delta \hat S^{(2)} &=\frac{1}{2}\sum_{\mu,\nu=0}^3\frac{\partial^2 \hat S}{\partial p_{\mu} \partial p_{\nu}}\delta
  p_{\mu}^{(1)}\delta p_{\nu}^{(1)} + \sum_{\mu=0}^3\frac{\partial \hat S}{\partial p_{\mu}}\delta p_{\mu}^{(2)} \,.
  \label{eq:deltap12}
\end{align}
The solutions of the above equations read
\begin{widetext}
  \begin{align}
    \delta\bm p^{(1)}=& \frac{1}{2}
    \begin{pmatrix}
      \csc\theta(r_2 \sin\gamma+r_1 \cos\gamma)\\
      r_2 \cos\gamma-r_1\sin\gamma\\
      r_3-\cot x (r_2\sin\gamma+r_1\cos\gamma)\\
      2 r_0\\
    \end{pmatrix},\\
    \delta\bm p^{(2)}=&\frac{1}{8}
    \begin{pmatrix}
      -\csc\theta (r_2 \cos\gamma-r_1 \sin\gamma)
      (2\cot\theta(r_2\sin\gamma+r_1\cos\gamma)-r_3)\\
      (r_2\sin\gamma+r_1\cos\gamma)
      (r_1\cos\gamma\cot\theta+r_2\sin\gamma \cot\theta-r_3)\\
      (r_2
      \cos\gamma-r_1 \sin\gamma) \left( (\cos (2 \theta)+3) \csc^2\theta (r_2
      \sin\gamma+r_1\cos\gamma)-2 r_3 \cot\theta\right)/2\\
      8k\delta L
    \end{pmatrix} \,.
    \label{eq:deltapsol}
  \end{align}
These equations define the Langevin process for the parameters $\bm p$. To
obtain the corresponding Fokker-Planck equation, we need to calculate the
average of $\delta \bm p^{(2)}$ and the (co)variance of $\delta \bm p^{(1)}$.
With the help of Eq.\ (\ref{eq:rmuvar}) we obtain
\begin{align}
  \label{eq:corr}
  \langle\delta\bm p^{(2)}\rangle&=
  \begin{pmatrix}
    0\\
    \tfrac{1}{2}\cot\theta\\
    0\\
    k
  \end{pmatrix}\delta L \,,\\
  \langle\delta\bm p^{(1)}\delta\bm p^{(1)T}\rangle&=
  \begin{pmatrix}
     \csc ^2\theta & 0 & -\cot\theta \csc\theta & 0 \\
      0 & 1 & 0 & 0 \\
     -\cot\theta \csc\theta & 0 & \csc^2\theta+\gamma-1 & 0 \\
      0 & 0 & 0 & \alpha  \\
  \end{pmatrix} \delta L \,.
\end{align}
\end{widetext}
Entering these correlators into the general form of the Fokker Planck
equation,~\cite{vankampen2007}
\begin{align}
  \frac{\partial P}{\partial L} = -\sum_{\mu=0}^3\partial_{p_{\mu}} \! \left( \frac{\langle\delta p_{\mu}^{(2)}\rangle}{\delta L} P
  \right)+\frac{1}{2}\sum_{\mu,\nu=0}^3\partial^2_{p_{\mu} p_{\nu}} \! \left( \frac{\langle\delta p_{\mu}^{(1)}\delta
  p_{\nu}^{(1)}\rangle}{\delta L} P\right),
\end{align}
we arrive at Eq.~(\ref{eq:FPeuler}) of the main text.

\bibliography{refs}

\end{document}